\documentclass[a4paper,11pt]{article}
\usepackage{jheppub} 
\usepackage{lineno}
\usepackage{amssymb}

\title{Generalized CV Conjecture and Krylov Complexity in Two-Mode Hermitian Systems via Information Geometry}

\author[1]{Ke-Hong Zhai,}
\author[1]{Lei-Hua Liu,}
\affiliation{1. Department of Physics, College of Physics, Mechanical and Electrical Engineering, Jishou University, Jishou 416000, China }

\emailAdd{2022700450@stu.jsu.edu.cn, liuleihua8899@hotmail.com, hqzhang@buaa.edu.cn}
\author[2,3]{Hai-Qing Zhang}
\affiliation{2. Center for Gravitational Physics, Department of Space Science, Beihang University, Beijing
100191, China}
\affiliation{3. Peng Huanwu Collaborative Center for Research and Education, Beihang University, Beijing 100191, China}

\begin{abstract}
{We extend the “complexity=volume” (CV) conjecture \cite{Zhang:1990fy} to quantum states of two-mode Hermitian systems using the framework of information geometry. Specifically, we conjecture that the Krylov complexity of a quantum state equals the volume of the Fubini-Study metric. To test this conjecture, we construct the wave functions for both closed and open two-mode systems. For the closed system, the wave function corresponds to the well-known two-mode squeezed state, while for the open system, we employ the second kind of Meixner polynomials to generate an open two-mode squeezed state. Remarkably, in both cases, the calculated Fubini-Study volume matches the Krylov complexity, providing analytic evidence for the generalized CV relation in this controlled two-mode setting. Our results establish a direct link between operator growth in Krylov space and geometric properties of quantum states, highlighting the potential applications of this framework in quantum information and quantum optics.}
\end{abstract}

\begin{document}
\maketitle
\flushbottom

\section{Introduction}
\label{sec:intro}

Recently, complexity has become a vital concept to describe the chaotic behaviors of a quantum system \cite{Nielsen:2006cea,Parker:2018yvk} or a spacetime \cite{Susskind:2014rva,Stanford:2014jda}. The concept of complexity was originally proposed to study the difficulty of a quantum system which can transform from one state to another. Therefore, it provides a new insight to probe the chaotic behaviors of a quantum system. In the spacetime, it was suggested that the complexity of a quantum state on the boundary is dual to the volume of the Einstein-Rosen (ER) bridge, which is the so-called CV conjecture \cite{Susskind:2014rva,Stanford:2014jda}. This kind of duality is one of the manifestations of holography \cite{Maldacena:1997re}. 

As a matter of fact, there are various kinds of complexity. For instance, in the quantum computings, Nielson {\it et.al.} have introduced the notion of  quantum "circuit complexity" to evaluate the minimum numbers of logical steps which is required to perform a task \cite{Nielsen:2006cea,Nielsen:2005mkt,Dowling:2006tnk}. Interestingly, this kind of quantum circuit complexity is closed related to the minimal length of paths in a curved geometry, which is a sign of the correspondence between the complexity and certain quantities in geometry.  Another method for calculating complexity of states in quantum field theory is based on the ``Fubini-Study" metric from the information geometry \cite{Chapman:2017rqy}. It should be noted that both of the above approaches depend on the choice of the parametric manifolds.

 Compared to the above two approaches, the Krylov complexity is free of ambiguities without needing the parametric manifolds in light of its definition \cite{Parker:2018yvk}. Krylov complexity was established to explore the growth of operators in quantum systems and to distinguish between the chaotic and integrable systems. Due to its uniqueness, Ref. \cite{Parker:2018yvk} proposed a universal bound for the Krylov complexity according to the specific Lanczos coefficients. Later on, the concept of Krylov complexity has been extended into a large number of fields, for instance in: the Sachdev-Ye-Kitaev (SYK) model \cite{Rabinovici:2020ryf,He:2022ryk}, the generalized coherent state \cite{Patramanis:2021lkx}, the Ising and Heisenberg
models \cite{Cao:2020zls,Trigueros:2021rwj,Heveling:2022hth}, conformal field theories \cite{Dymarsky:2021bjq,Caputa:2021ori}, topological phases of matter \cite{Caputa:2022eye}, integrable models with
saddle-point dominated scrambling \cite{Bhattacharjee:2022vlt}, cosmological complexity \cite{Adhikari:2022oxr}, thermal quantum field theories \cite{Camargo:2022rnt,He:2024hkw,He:2024xjp} and $\it e.t.c$. The very recent developments can be found in Refs. \cite{Hashimoto:2023swv,Vasli:2023syq,Gill:2023umm,Bhattacharjee:2023uwx,Adhikari:2022whf,Caputa:2024vrn,Basu:2024tgg,Sasaki:2024puk,Caputa:2024xkp,Sahu:2024opm,Bhattacharjee:2022qjw,Kim:2021okd,Chen:2024imd,Bhattacharjee:2024yxj,Sanchez-Garrido:2024pcy,Chattopadhyay:2024pdj,Balasubramanian:2024ghv,Bhattacharya:2024hto,Mohan:2023btr,He:2024pox,Bhattacharya:2022gbz,Bhattacharya:2023zqt}. And there is a very comprehensive review of Krylov complexity \cite{Nandy:2024htc}. In addition to these developments in Krylov complexity, two-mode and multimode Hermitian Hamiltonian systems have also appeared in a wide range of physical contexts. For example, correlated two-mode fields and deformed atom-field interactions have been used to study entanglement, Fisher information and parameter estimation in quantum optics and quantum information \cite{Berrada:2022tzi}. Related Hamiltonian structures also arise in generalized or fractional quantum-mechanical systems, where the wave functions and probability distributions are modified by nonstandard kinetic or interaction terms \cite{al2021applying}. Furthermore, Hermitian and pseudo-Hermitian descriptions have been employed in quantum-emitter systems and open photonic structures \cite{li2022coupling}. These studies indicate that two-mode or multimode Hamiltonian structures provide useful theoretical laboratories for exploring quantum correlations, information geometry and dynamical complexity beyond simple single-mode models.

There have been several efforts to incorporate Krylov complexity into the framework of holography. For instance, Ref. \cite{Rabinovici:2023yex} established a one-to-one correspondence between the thermo-field double (TFD) state (at infinite temperature) and Jackiw-Teitelboim (JT) gravity  \cite{Jackiw:1984je,Teitelboim:1983ux}. Then, the Krylov complexity can be computed in the SYK model using JT gravity through holography. Additionally, Ref. \cite{Kar:2021nbm} evaluated the evolution of Krylov complexity in JT gravity using the random matrix theory. It is well-known that the information geometry connects geometry to quantum states through the Fisher metric (or Fubini-Study metric) \cite{2018arXiv180808271N}. Therefore, from this perspective, Ref. \cite{Caputa:2021sib} systematically investigated the geometry of Krylov complexity and concluded that \( V_t = 2\pi K_{\mathcal{O}} \) where $V_t$ is the volume of the Fubini-Study metric and $K_{\mathcal{O}}$ is the Krylov complexity of the operator $\mathcal{O}$. However, they only investigated the closed system with single mode. 

In this work, we propose the generalized CV conjecture in the closed and open systems from the Hermitian Hamiltonian. In order to test our conjecture, we study the two-mode Hermitian Hamiltonian. For the closed system, the wave function is constructed from the two-mode squeezed state by employing the displacement operators. As for the open system, the construction of the wave functions is more involved, which needs the second kind of Meixner polynomials \cite{Zhai:2024odw,Parker:2018yvk}. The resulting wave function is dubbed open two-mode squeezed state. Eventually, we find that the relation $V_t=2\pi K_O$ holds for the closed and open two-mode Hermitian systems considered here, providing analytic evidence for the generalized CV relation within this controlled setting.

This paper will be organized as follows. In Sec.\ref{section Lanczos algorithm}, we briefly introduce the Lanczos algorithm in closed system and open system, respectively. In Sec. \ref{wave function}, the wave function for the closed system and open system will be derived. Sec. \ref{generalized cv conjecture} will give the generalized CV conjecture. In Sec. \ref{section conclusions}, the conclusion and outlook will be given.

\section{Lanczos Algorithm }
\label{section Lanczos algorithm}
In this section, we will briefly introduce the Lanczos algorithm for the closed system and open system, respectively. All of the quantum operators are defined in the Heisenberg picture
\begin{equation}
    \partial_t \mathcal{O}(t)=i[H,\mathcal{O}(t)],
    \label{operator O}
\end{equation}
where $H$ is the Hamilton and its corresponding solution can be written by 
\begin{equation}
    \mathcal{O}(t)=e^{iHt}\mathcal{O}e^{-iHt}.
    \label{solution of O}
\end{equation}
\subsection{For the closed system}
In the framework, the most important operator is the Liouvillian super-operator $\mathcal{L}_X$ as $\mathcal{L}_X Y=[X,Y]$.  We apply this to Eq. \eqref{solution of O}, the operator can also be expressed as  
\begin{equation}
    \mathcal{O}(t)=e^{i\mathcal{L}t}\mathcal{O}=\sum_{n=0}^{\infty} \frac{(it)^n}{n!}\mathcal{L}^{n}\mathcal{O}(0)=\sum_{n=0}^{\infty} \frac{(it)^n}{n!}\Tilde{\mathcal{O}}_{n}.
    \label{O with L}
\end{equation}
we will use eq. \eqref{O with L}.

Furthermore, we could define the $\mathcal{L}^{n}\mathcal{O}=\Tilde{\mathcal{O}}_{n}=[H,\Tilde{O}_{n-1}]$, where it serves as the basis of Hilbert space, 
\begin{equation}
\begin{split}
\mathcal{O}\equiv |{\tilde{\mathcal{O}} })  ,\mathcal{L}^{1}\mathcal{O}\equiv |{\tilde{\mathcal{O}} }_{1})  ,\mathcal{L}^{2}\mathcal{O}\equiv |{\tilde{\mathcal{O}} }_{2})  ,\mathcal{L}^{3}\mathcal{O}\equiv |{\tilde{\mathcal{O} } }_{3}) ...   \end{split}
\label{basis of O}
\end{equation}
However, they are not orthogonal bases. We could utilize the so-called Lanczos algorithm to construct the orthogonal basis, the first two operators become
\begin{equation}
    \mathcal{O}_{0}=|\Tilde{\mathcal{O}}_{0})=\mathcal{O}, ~~~|\mathcal{O}_{1})=b_{1}^{-1}\mathcal{L}|\mathcal{O}_{0})
    \label{first two O}
\end{equation}
where $b_{1}=\sqrt{(\Tilde{\mathcal{O}}_{0}\mathcal{L}|\mathcal{L}\Tilde{\mathcal{O}}_{0})}$ describes the normalized vector. The higher order orthogonal basis is written as 
\begin{equation}
    |\mathcal{O}_{n})=b_{n}^{-1}|A_{n})
    \label{n-th turns O}
\end{equation}
with
\begin{equation}
    |A_{n})=\mathcal{L}|\mathcal{O}_{n-1})-b_{n-1}|\mathcal{O}_{n-2}),~~~b_{n}=\sqrt{(A_{n}|A_{n})}
    \label{n-th A}
\end{equation}
where $b_{n}$ denotes the Lanczos coefficient, which plays a crucial role in determining the nature of the dynamical system, such as whether it is chaotic, integrable, or free. For further details on the Lanczos coefficient, please refer to Ref. \cite{Parker_2019}. This iterative relation \eqref{n-th A} will stop until it produces the finite orthogonal Krylov basis if $b_{n}=0$. Subsequently, we can express Eq. \eqref{O with L} as follows,
\begin{equation}
    \mathcal{O}(t)=e^{i\mathcal{L}t}\mathcal{O}=\sum_{n=0}^{\infty} (i)^{n}\phi_{n}(t)|\mathcal{O}_{n})
    \label{O with O_n}
\end{equation}
where $\phi_{n}$ is the amplitude of the wave function satisfying with $\sum_{n} |\phi_{n}|^{2}=1$. As discussed in \cite{Parker_2019}, the Liouvillian superoperator \(\mathcal{L}\) can be interpreted as the Hamiltonian in terms of creation and annihilation operators. We already know the Lanczos algorithm, essentially it could construct the following recurrence relation due to the Gram-Schmidt method, 
\begin{equation} \mathcal{L}|\mathcal{O}_n)=b_{n+1}|\mathcal{O}_{n+1})+b_n|\mathcal{O}_{n-1}). 
    \label{eq L in K-basis of closed system}
\end{equation}
By plugging Eq. \eqref{O with O_n} into the Schr$\Ddot{o}$dinger equation, we can explicitly obtain
\begin{equation}
    \partial_{t}\phi_{n}(t)=b_{n}\phi_{n-1}-b_{n+1}\phi_{n+1}. 
    \label{O in SE}
\end{equation}
We should also note that Krylov complexity is generally difficult to measure directly. Its extraction requires the reconstruction of the Krylov basis and the Lanczos coefficients from operator dynamics or dynamical correlation functions. In many-body or experimental systems, this procedure can be limited by the rapid growth of Krylov space, finite measurement precision, noise and the difficulty of accessing high-order correlation functions. From a computational perspective, repeated applications of the Liouvillian operator and the Gram-Schmidt orthogonalization procedure become increasingly expensive as the effective Krylov dimension grows. The present two-mode model is analytically tractable because the Lanczos recursion can be mapped to an orthogonal-polynomial problem, while the practical computation of Krylov complexity in generic many-body systems remains a challenging task.

After these general remarks, we now focus on the closed-system case, where the Liouvillian superoperator can be written in terms of creation and annihilation operators as 
\begin{equation}
    \mathcal{L}=\alpha(\mathcal{L}_-+\mathcal{L}_+),
    \label{general form for closed system}
\end{equation}
where $\mathcal{L}_-$ represents the part of annihilation operator and $\mathcal{L}_+$ denotes the part of creation operator \cite{Caputa:2021sib}. However, the real Hamiltonian is more complicated, including the extra part denotes the part of open system. 

\subsection{For the open system}
\label{for the open system}
For depicting the open system, we focus on the generalized Lanczos algorithm \cite{Bhattacharjee:2022lzy}, which is derived from the Lindblad master equation \cite{Gorini:1975nb}. The generalized recurrence relation associated with this algorithm can be expressed as follows:
\begin{equation}
    \mathcal{L}|\mathcal{O}_n)=-ic_n|\mathcal{O}_n)+b_{n+1}|\mathcal{O}_{n+1})+b_n|\mathcal{O}_{n-1}),
    \label{eq L in K-basis}
\end{equation}
where we will follow the notation of our previous work \cite{Zhai:2024odw, Li:2024kfm} and $c_n$ encodes the information of the open system for a certain Hamiltonian. We followed a specific notation from the perspective of the matrix representation of quantum mechanics. Here, \( ic_n \) arises from the diagonal component of the matrix. Based on the work of \cite{Bhattacharjee:2022lzy}, we observed that the \(\mathcal{O}_n\) part corresponds to the open system, while \(\mathcal{O}_{n+1}\) and \(\mathcal{O}_{n-1}\) relate to the closed system. From this perspective, the generalized Lanczos algorithm can extract information from a specific Hamiltonian encoded in \( ic_n \), similar to our earlier studies \cite{Li:2024ljz, Li:2024iji, Li:2024kfm, Zhai:2024odw}. More specifically, once the Hamiltonian is established, the generalized Lanczos algorithm \eqref{eq L in K-basis} can be employed to distinguish which components represent an open system and which parts represent a closed system. Reference \cite{Li:2024kfm} even discusses the possibility that a Hermitian Hamiltonian could exhibit characteristics of an open system.

 Let us discuss more about the $c_n$. Recalling that the Eq. \eqref{O in SE} (Schr$\Ddot{o}$dinger equation essentially) only includes the Lanczos coefficient. The parameter $c_n$ will appear in the following modified equation,
\begin{equation}
    \partial_\eta\phi+2b_n\partial_n\phi+\tilde{c}_n\phi=0,
    \label{equ of cn}
\end{equation}
where $\tilde{c}_n=-ic_n$ is defined, meanwhile we have utilized the $b_{n+1}\approx b_n$ and $\phi_{n+1}-\phi_{n-1}\approx 2\partial_n \phi$ in the continuous limit. From Ref. \cite{Parker:2018yvk}, the generic formulas of $b_n$ and $c_n$ are
\begin{equation}
    b_{n}^{2}=|1-u_{1}^{2}|n(n-1+\beta),\ \ c_{n}=i u_{2}(2n+\beta)
    \label{eq bn and cn}
\end{equation}
where $u_1$, $u_2$ and $\beta$ are determined by various models and $u_2$ plays a role of dissipative coefficient also shown in Ref. \cite{Li:2024kfm}. Particularly, $b_n\propto \alpha n$ indicates that the system is a maximal chaotic system. And the dynamical system is a weak dissipative system as $u_2\ll 1$, which could be approximated to be a closed system as mentioned in \cite{Li:2024ljz, Li:2024iji, Li:2024kfm, Zhai:2024odw}.

\section{Wave function}
\label{wave function}
The essence of the generalized Lanczos algorithm \eqref{eq L in K-basis} is a recurrence relation belonging to the second kind of Meixner polynomials \cite{viswanath1994recursion,Hetyei_2009}. 
Equation \eqref{eq L in K-basis} can be rewritten in the form of an orthogonal polynomial sequence (OPS):
\begin{equation}
    P_{n+1}(x)=(x-\tilde{c}_n)P_n(x)-b_n^2P_{n-1}(x),
    \label{eq definition Pn}
\end{equation}
where we define $\tilde{c}_n=-ic_n$ in Eq. \eqref{equ of cn}, and $x$ represents the Hamiltonian. It is useful to clarify the mathematical role of the Meixner polynomials of the second kind in our construction. The recurrence relation (3.1) belongs to an orthogonal polynomial sequence, where the coefficients $b_n$ and $\tilde c_n$ determine the corresponding Jacobi matrix representation of the Liouvillian operator. The orthogonality of these polynomials ensures that the Krylov basis generated by the generalized Lanczos algorithm is well defined. In this sense, the Meixner polynomials provide a bridge between the tridiagonal Lanczos representation and the wave function in Krylov space. Their generating function allows us to resum the Krylov expansion and obtain the open two-mode squeezed state in Eq. (3.3). Since the focus of this work is the generalized CV relation rather than the general theory of orthogonal polynomials, we only use the properties that are directly relevant to the construction of the Krylov wave function. The sequence begins with the initial conditions $P_0(x) =1$ and $P_1(x) = x-c_0$. The polynomial $P_n(x)=\det (x-\mathcal{L}_n)$ is for an open system, and $\mathcal{L}_n$ is the Liouvillian superoperator for the $n$-th quantum state. By introducing the natural orthonormal basis $\{e_n\}$. We could represent $ |P_{n}(x))=\Bigl (\prod_{k=1}^{n}b_{k}\Bigr )|\mathcal{O}_n),\ \ \mbox{and}\ \ \ |x^{n})=\mathcal{L}^{n}|\mathcal{O}).$  Combining Eq. \eqref{eq definition Pn}, there is an explicit relation as follows,
\begin{equation}
    b_{n+1}e_{n+1}+b_ne_{n-1}=(x-\tilde{c}_n)e_n,
    \label{equalled relation of bn}
\end{equation}
where $e_n=\mathcal{O}_n$ in our case and it is equivalent to the generalized Lanczos algorithm \eqref{eq L in K-basis}. Before deriving the explicit wave function, let us clarify the role of the initial condition. 
In the Krylov construction, the initial state is chosen as the first normalized Krylov basis 
vector $|e_0)$, which is generated from the initial operator $O$. This choice corresponds 
to the standard initialization of the Lanczos algorithm and fixes the normalization of the 
Krylov wave function. In the two-mode realization considered here, this initial Krylov state 
is mapped to the two-mode vacuum state $|0;0\rangle_{\vec k,-\vec k}$. The subsequent 
evolution generated by the Liouvillian then produces the Krylov amplitudes along the 
basis $|e_n)$, which are identified with the two-mode number states $|n;n\rangle_{\vec k,-\vec k}$. 
Different initial operators may lead to different Krylov bases and different wave functions. 
In the present work, however, the choice of $|e_0)$ provides the minimal and analytically 
tractable initial condition for testing the generalized CV relation. Being armed with generating function of Meixner polynomials, the corresponding wave function can be derived by (the details can be found in Ref. \cite{Zhai:2024odw})
\begin{eqnarray}
    |\mathcal{O}(\eta))=e^{i\mathcal{L}\eta}|e_{0})
    =\frac{{\rm sech}~  r_k}{1+u_2\tanh r_k}\sum_{n=0}^{\infty}|1-u_1^{2}|^{\frac{n}{2}}\frac{(-\exp(2i\phi_k)\tanh r_k)^{n}}{(1+u_2\tanh r_k)^{n}}|e_{n}).
    \label{open two mode state}
\end{eqnarray}
where $r_k\equiv r_k(\eta)$, $\phi_k\equiv\phi_k(\eta)$ are the parameters depicting the feature of two-mode squeezed state (more details can be found in Refs. \cite{Li:2023ekd,Li:2021kfq,Liu:2021nzx}) and $|e_n)=|n;n\rangle_{\vec{k},-\vec{k}}$ in this work (utilized for testing our conjecture in the later investigations). The parameters $u_1$ and $u_2$ are determined by Hamiltonian defined in \eqref{eq bn and cn}. This wave function is dubbed as the open two-mode squeezed state since the leading order in terms of $u_2$ is exactly the two-mode squeezed state
\begin{eqnarray}
  |\mathcal{O}(\eta))= \hat{S}_{\Vec{k}}(r_{k},\phi_{k})|0;0\rangle_{\vec{k},-\vec{k}}
  =\frac{1}{\cosh r_k}\sum_{n=0}^{\infty} (-1)^ne^{2in\phi_k}\tanh^nr_k|n;n\rangle_{\vec{k},-\vec{k}}, 
  \label{two mode squeezed state}
\end{eqnarray}  
$\hat{S}_{\vec{k}}$ plays a role of displacement operator and its definition can be found in Eq. \eqref{displacment operator of two mode squeezed state}.

The derivation of this wave function is highly significant as it is model-independent, relying only on the recurrence relation \eqref{eq definition Pn}. Information of various models is encoded in the parameters \(u_1\) and \(u_2\), with \(u_2\) serving as a dissipative coefficient, as noted in Refs. \cite{Li:2024kfm, Li:2024iji, Zhai:2024odw}. Although our wave function is derived within the framework of cosmological perturbation theory for the inflation, it can also be applied to fields such as condensed matter, quantum information, quantum optics, and $\it etc$ \cite{Nandy:2024htc}.

\section{The generalized CV conjecture}
\label{generalized cv conjecture}
The CV conjecture arises from the analysis of the complexity of quantum states and is related to the ER bridge within the holographic framework \cite{Stanford:2014jda}. It is reasonable to extend this conjecture to broader quantum systems. In \cite{Caputa:2021sib}, the geometry of Krylov complexity is examined using the Fubini-Study metric. The authors concluded that \( V_g = 2\pi \mathcal{C}_{K} \), where \( V_g \) represents the volume of the Fubini-Study metric and \( \mathcal{C}_K \) denotes the Krylov complexity. However, their findings are applicable only to closed systems, since the Liouvillian superoperator only contributes to the Lanczos coefficient part ($b_n$ part) with single mode whose Hamiltonian is of the form \eqref{general form for closed system}. Nevertheless, Ref. \cite{Caputa:2021sib} provides a strong hint that we could extend the CV conjecture within the framework of information geometry, as it explicitly connects probability to Riemannian geometry. Before testing the generalized CV relation, we briefly review the role of the Fubini-Study metric in quantum-state geometry.

\subsection{Fubini-Study Metric and Its Implications}

The Fubini-Study metric provides a natural Riemannian metric on the projective Hilbert space of quantum states, allowing a geometric interpretation of quantum evolution \cite{Nielsen:2006cea,Nielsen:2005mkt,Dowling:2006tnk,2018arXiv180808271N,Caputa:2021sib}. For two infinitesimally close normalized states $|\psi\rangle$ and $|\psi + d\psi\rangle$, the line element is defined as
\begin{equation}
ds^2_{FS} = \frac{\langle d\psi | d\psi \rangle}{\langle \psi | \psi \rangle} - \frac{|\langle \psi | d\psi \rangle|^2}{\langle \psi | \psi \rangle^2},
\end{equation}
which quantifies the ``distance'' between quantum states in Hilbert space. Physically, this metric captures the distinguishability between nearby states and serves as a measure of quantum coherence and fluctuations.

In the context of quantum complexity, the Fubini-Study metric has been employed to define the minimal length of paths connecting quantum states, corresponding to the minimal number of unitary operations required for a given quantum evolution \cite{Nielsen:2006cea}. More recently, this geometric perspective has been related to Krylov complexity: the volume of the Fubini-Study metric associated with a quantum state provides a measure of operator growth and chaotic behavior \cite{Caputa:2021sib}. 

In our study, we utilize this framework to relate the Krylov complexity of two-mode Hermitian systems to the Fubini-Study volume. Specifically, given a state $|z\rangle = |O_n\rangle$, the metric can be written as
\begin{equation}
ds^2 = \frac{\langle dz | dz \rangle}{\langle z | z \rangle} - \frac{\langle dz | z \rangle \langle z | dz \rangle}{\langle z | z \rangle^2},
\label{fubuni studu metric}
\end{equation}
which recovers the standard Fubini-Study form in the complex coordinates $(r, \varphi)$ for two-mode squeezed states. It is important to clarify the meaning of the volume appearing in our generalized CV relation. In the present work, $V_t$ does not denote a spacetime volume or the volume of an Einstein-Rosen bridge. Instead, it is the Riemannian volume induced by the Fubini-Study metric on the parameter manifold of the Krylov wave function. For the two-mode squeezed state considered here, this parameter manifold is described by the coordinates $(r,\varphi)$, and the volume is defined as
\begin{equation}
V_t=\int_{\mathcal{M}_t} d r\, d\varphi\,\sqrt{g},
\end{equation}
where $\mathcal{M}_t$ denotes the region $0\leq r\leq r_k(t)$ and $0\leq\varphi<2\pi$. Therefore, the equality $V_t=2\pi K_O$ should be understood as a relation between Krylov complexity and the information-geometric volume of the associated state manifold. This formulation establishes a direct connection between operator growth in Krylov space and the geometric properties of quantum states, providing a solid foundation for testing the generalized CV conjecture.

\subsection{Limitations and Assumptions of the Information-Geometric Framework}

Although the Fubini-Study metric provides a natural geometric structure on the projective Hilbert space, its application to Krylov complexity relies on several assumptions. First, the standard Fubini-Study metric is most naturally defined for normalized pure states generated by unitary evolution. Therefore, our construction is directly applicable to Hermitian Hamiltonians, for which the inner-product structure of the Hilbert space is preserved. For genuinely non-Hermitian systems or general Lindblad dynamics, the time evolution is usually non-unitary, and the usual Fubini-Study metric may not be sufficient. In such cases, one may need to employ biorthogonal quantum geometry, the Bures metric, or the quantum Fisher information metric.

Second, the Weyl or two-photon algebra used in this work provides a convenient and analytically tractable framework for quadratic bosonic Hamiltonians and squeezed-state dynamics. However, this algebraic structure does not describe all possible quantum systems. Strongly interacting many-body systems, spin systems, non-Gaussian states, or Hamiltonians with higher-order nonlinear interactions may require different algebraic structures and may lead to Krylov wave functions different from the two-mode squeezed-state form considered here.

Third, the information-geometric volume depends on the chosen parameter manifold. In the present work, the coordinates $(r,\varphi)$ are physically motivated by the two-mode squeezed state and the open two-mode squeezed state. This choice allows us to obtain a two-dimensional Fubini-Study metric and to compare its volume with Krylov complexity. For more general quantum states, the parameter space may become higher-dimensional, or may not admit a simple global parametrization. In such cases, the definition and interpretation of the corresponding volume can become more subtle.

Therefore, the relation $V_t=2\pi K_O$ should be understood as analytic evidence for the class of Hermitian two-mode Hamiltonians considered here, rather than as a universal identity for arbitrary quantum systems. The main assumptions behind our conjecture are Hermiticity, the two-mode quadratic algebraic structure, the use of pure-state Fubini-Study geometry, and the existence of a well-defined Krylov basis generated by the Lanczos recursion. Extensions to non-Hermitian systems, mixed states, strongly interacting models, and higher-dimensional parameter manifolds remain important directions for future investigation.

\subsection{Generalized CV conjecture for closed system}

To test this generalized CV conjecture, we will first focus on the part of closed system. In particular, we will examine the two-mode Hermitian Hamiltonian, whose group representation is the Weyl algebra. Let us recall that the methology of \cite{Caputa:2021sib}, where the key ingredient is the generalized displacement operator according to the Hamiltonian. For various groups, its representation theory is different. The creation and annihilation operators of every group are also distinctive, leading the different wave functions.

In light of this logic, we see that the excited state of the wave function for the Weyl algebra correspond to the well-known two-mode squeezed state. In this sense, the two-mode squeezing operator can be regarded as a generalized displacement operator associated with the quadratic generators of the two-mode algebra. We will continue applying this method for the two-mode system. First, we can write down the displacement operator as
\begin{equation}
    D(\eta)=S(\eta)=e^{iHt}=\exp(-\eta\hat{a}_{-\vec{k}}^{\dagger}\hat{a}_{\vec{k}}^{\dagger}+\bar{\eta}\hat{a}_{\vec{k}}\hat{a}_{-\vec{k}}),
    \label{displacment operator of two mode squeezed state}
\end{equation}
where $\eta$ is a complex number, $\bar{\eta}$ is the complex conjugate of $\eta$ and $t$ is a real parameter. It should be noted that the operator in Eq.~\eqref{displacment operator of two mode squeezed state} is a generalized displacement operator in the sense of coherent-state theory. For the ordinary Weyl-Heisenberg algebra, the displacement operator generates coherent states through linear combinations of $\hat a^\dagger$ and $\hat a$. In the present two-mode case, however, the relevant generators are quadratic operators, $\hat a^\dagger_{\vec k}\hat a^\dagger_{-\vec k}$ and $\hat a_{\vec k}\hat a_{-\vec k}$, so that the corresponding generalized displacement operator becomes the two-mode squeezing operator. Therefore, it belongs to the class of Gaussian unitary operations, together with phase rotations and beam-splitter transformations. While phase rotations change the phases of the modes and beam-splitter operations mix different modes, the two-mode squeezing operation creates correlated pairs and generates entanglement between the modes. This is why it naturally produces the two-mode squeezed state used in our Krylov construction.

Then, using the $ \hat{a}^{\dagger}|n\rangle=\sqrt{n+1}|n+1\rangle,\ \hat{a}|n\rangle=\sqrt{n}|n-1\rangle$, and changing the variable $\eta=re^{2i\phi}$, we can explicitly obtain the two-mode squeezed state \eqref{two mode squeezed state}. According to the definition of Krylov complexity $ K_{\mathcal{O}}=\sum_{n}n|\varphi_n(t)|^2$, we could obtain the Krylov complexity as,
\begin{equation}
   K_{\mathcal{O}}=\sinh^2r_k.
    \label{eq K-complexity}
\end{equation}
To relate CV conjecture, we need Fubini-Study metric 
\eqref{fubuni studu metric}. Under the complex coordinates $(r,\phi)$, the metric of two-mode squeezed state can be explicitly obtained as follows,
\begin{equation}
    ds^2=\frac{dzd\bar{z}}{(1-z\bar{z})^2}=dr^2+\sinh^22rd\phi^2.
    \label{metric of two mode}
\end{equation}
Consequently, one can easily obtain the corresponding volume of this metric, 
\begin{equation}
    V_t=\int_{0}^{r_k}dr\int_{0}^{2\pi}d\phi\sqrt{g}=2\pi\sinh^2 r_k=2\pi K_{\mathcal{O}},
    \label{volume of two mode}
\end{equation}
where $r_k$ should be positive according to the physical meaning of two-mode squeezed state. Thus, we could see that the conjecture $V_t=2\pi K_{\mathcal{O}}$ is valid for the closed system within two-mode Hermitian Hamiltonian. It is also worth noting that the two-mode squeezed state is a paradigmatic entangled state. The squeezing parameter $r_k$ controls both the mean occupation number and the correlation strength between the two modes. Therefore, the result $K_O=\sinh^2 r_k$ indicates that the Krylov complexity grows together with the two-mode correlations. From this perspective, the Fubini-Study volume not only measures the geometric growth of the Krylov wave function, but also encodes part of the entanglement structure of the two-mode state.

\subsection{Generalized CV conjecture in open system }
\label{CV in open system}

The Hermitian Hamiltonian of an open system can take various forms. Ref. \cite{zhang1990coherent} has systematically demonstrated that two-mode squeezed states belong to the ``two-photon" algebra, which is spanned by specific operators
\begin{equation}
    \hat{a}_i^{\dagger}\hat{a}_j^{\dagger},\ \ \ \hat{a}_i^{\dagger}\hat{a}_j+\frac{1}{2}\delta_{ij},\ \ \ \hat{a}_i\hat{a}_j,\ \ \ \hat{a}_i^{\dagger},\ \ \ \hat{a}_i,\ \ \ I,
    \label{general hamiltonian with two mode}
\end{equation}
where \(i, j = \vec{k}, -\vec{k}\) represent the two modes. The construction of the Hamiltonian is constrained by several symmetries. First, Hermiticity requires that the pair-creation term and the pair-annihilation term appear 
with complex-conjugate coefficients. Second, the two-mode squeezed state naturally pairs the modes $\vec k$ and $-\vec k$, so the Hamiltonian should preserve this mode-pair structure. Third, the diagonal number-operator sector preserves the occupation-number 
basis and corresponds to the $c_n$ component in the generalized Lanczos recurrence. These symmetry considerations restrict the allowed quadratic operators and motivate the Hamiltonian used below. Based on these considerations, the general Hermitian Hamiltonian with two modes can be expressed as:
\begin{equation}
    H=u_2(\hat{a}_{\vec{k}}^{\dagger}\vec{a}_{\vec{k}}+\vec{a}_{-\vec{k}}\hat{a}_{-{\vec{k}}}^{\dagger})+i(\xi \hat{a}_{-\vec{k}}^{\dagger}\hat{a}_{\vec{k}}^{\dagger}-\bar{\xi}\hat{a}_{\vec{k}}\hat{a}_{-\vec{k}}),
    \label{general open two mode hamiltonian}
\end{equation}
where $\xi$ is a complex parameter controlling the strength and phase of the pair-creation and pair-annihilation processes, with $\bar{\xi}$ denoting its complex conjugate. The parameter $u_2$ is associated with the diagonal number-operator sector and corresponds to the coefficient $c_n$ in the generalized Lanczos recurrence relation. In this sense, $u_2$ encodes the open-system or dissipative contribution to the Krylov dynamics. The parameter $u_1$, which appears in the Lanczos coefficient $b_n$ in Eq.~\eqref{eq bn and cn}, controls the off-diagonal hopping in Krylov space. Therefore, the parameters in the Hamiltonian are not arbitrary: they are fixed by the Hermiticity condition, the two-mode algebraic structure, and the generalized Lanczos coefficients $(b_n,c_n)$. To relate Eq. \eqref{displacment operator of two mode squeezed state}, we could obtain $\eta=\xi t$ and $\bar{\eta}=\bar{\xi}t$. This Hamiltonian indicates that it is impossible to define the generalized displacement operator due to the presence of the term $u_2(\hat{a}_{\vec{k}}^{\dagger}\vec{a}_{\vec{k}}+\vec{a}_{-\vec{k}}\hat{a}_{-{\vec{k}}}^{\dagger})$. To construct the corresponding wave function, we need to utilize the second kind of Meixner polynomials, as discussed in \cite{Zhai:2024odw}. The resulting wave function is given by \eqref{open two mode state}. 

To test $V_t=2\pi K_{\mathcal{O}}$, we need to calculate the Fubini-Study metric. In light of the open two-mode squeezed state \eqref{open two mode state}, we could explicitly obtain the Krylov complexity as, 
\begin{equation}
    K_{\mathcal{O}}=\frac{|1-u_1^2|\tanh^2 t}{1+2u_2\tanh t+(u_2^2-|1-u_1^2|)\tanh^2t}.
    \label{open krylov complexity}
\end{equation}
where we have normalized the wave function for the Krylov complexity 
\begin{eqnarray}
   K_{\mathcal{O}}=\frac{1}{(\mathcal{O}(t)|\mathcal{O}(t))}\sum_nn|\varphi_n(t)|^2 .
   \label{normalized krylov complexity}
\end{eqnarray}
 The formula \eqref{open krylov complexity} is consistent with \cite{Bhattacharjee:2022lzy}. Next, we will use the Fubini-Study metric to evaluate the generalized CV conjecture. In our calculations, we will consider \( u_1 \) and \( u_2 \) as time-independent variables, determined by various models. For convenience, we define \( |z\rangle = |\mathcal{O}(\eta)) \). After some algebra, we can derive the corresponding Fubini-Study metric as follows (details can be found in Appendix \ref{appendix the calculation of the metric within the open system}):
\begin{eqnarray}
  ds^2&=&\Bigl (\frac{2}{\sinh 2r}-\frac{u_2\rm sech^2r}{1+u_2\tanh r}\Bigr )^2\frac{A}{(1-A)^2}dr^2+4\frac{A}{(1-A)^2}d\phi^2,   
\end{eqnarray}
with $A=|1-u_1^2|\tanh^2r/(1+u_2\tanh r)^2$ and we can obtain 
\begin{equation}
  \frac{A}{(1-A)^2}=\frac{|1-u_1^2|\tanh^2r(1+u_2\tanh r)^2}{[1+2u_2\tanh r+(u_2^2-|1-u_1^2|)\tanh^2r]^2}. 
  \label{some formula}
\end{equation}
For the parameter region considered in this work, namely $r>0$ and 
$1+u_2\tanh r>0$, the positive volume element can be related to the 
radial derivative of the Krylov complexity. Defining
\begin{equation}
K_O(r)=
\frac{|1-u_1^2|\tanh^2 r}
{1+2u_2\tanh r+(u_2^2-|1-u_1^2|)\tanh^2 r},
\end{equation}
one finds
\begin{equation}
\frac{dK_O(r)}{dr}
=
\frac{2A}{(1-A)^2}
\left(
\frac{2}{\sinh 2r}
-
\frac{u_2 \rm sech^2 r}{1+u_2\tanh r}
\right),
\end{equation}
where
\begin{equation}
A=
\frac{|1-u_1^2|\tanh^2 r}{(1+u_2\tanh r)^2}.
\end{equation}
Thus, in this parameter region,
\begin{equation}
\sqrt{g}=\frac{dK_O(r)}{dr}.
\end{equation}
Consequently, the Fubini-Study volume reduces to a boundary contribution at the upper limit of the radial coordinate.

With these formulae, we can obtain the square root of the metric determinant as, 
\begin{eqnarray}
    \sqrt{g}&=& \Bigl |\frac{2}{\sinh 2r}-\frac{u_2\rm sech^2r}{1+u_2\tanh r}\Bigr |\frac{2A}{(1-A)^2}
    = \Bigl |\frac{2}{\sinh 2r}-\frac{u_2\rm sech^2r}{1+u_2\tanh r}\Bigr |\times \nonumber\\
    &&\frac{2|1-u_1^2|\tanh^2r(1+u_2\tanh r)^2}{[1+2u_2\tanh r+(u_2^2-|1-u_1^2|)\tanh^2r]^2}.
    \label{determinant of g}
\end{eqnarray}
Remarkably, we find that the generalized conjecture still holds for the open system with Hermitian conditions, 
\begin{eqnarray}
    V_t=\int_{0}^{ t}dr\int_{0}^{2\pi}d\phi\sqrt{g}
    =2\pi\frac{|1-u_1^2|\tanh^2 t}{1+2u_2\tanh t+(u_2^2-|1-u_1^2|)\tanh^2t}=2\pi K_{\mathcal{O}}.
    \label{eq Vt and K}
\end{eqnarray}
Until now, we have tested the generalized CV conjecture for a general Hermitian two-mode Hamiltonian constructed from creation, annihilation, and number operators. These results suggest that the generalized CV relation may hold for a broader class of Hermitian Hamiltonians containing both closed- and open-system components. However, a complete proof for arbitrary Hermitian, multimode, strongly interacting, or non-Hermitian systems is still beyond the scope of the present work.

\section{Conclusions and outlook}
\label{section conclusions}
The work by \cite{Caputa:2021sib} has constructed wave functions using the generalized displacement operator, where the representation theory of distinct groups generates various displacement operators. Moreover, they discovered that the Krylov complexity of different quantum states is proportional to the volume of the corresponding Fubini-Study metric, although their analysis is limited to closed systems with a single mode. In light of this logic, we propose the generalized CV conjecture to quantum states generated by displacement operators within the framework of information geometry. Specifically, we test the relation $V_t=2\pi K_O$ for Hermitian two-mode Hamiltonians containing both closed- and open-system components. To test our conjecture, we examined the general two-mode Hermitian Hamiltonian, characterized by the Weyl algebra. For the closed part, the wave function is generated using the generalized displacement operator, as demonstrated in Eq. \eqref{displacment operator of two mode squeezed state}. Here, we can explicitly prove that \( V_t = 2\pi K_{\mathcal{O}} \), as shown in \eqref{volume of two mode}. However, when considering the open system component, the wave function cannot be derived using the approach from \cite{Caputa:2021sib}. Instead, we utilize the second kind of Meixner polynomials for construction, as indicated in Eq. \eqref{open two mode state}. Using the corresponding Fubini-Study metric, we find that the same relation 
$V_t=2\pi K_O$ continues to hold for the open two-mode squeezed state.

It should be emphasized that the volume appearing in the present work is the Fubini-Study volume of the Krylov wave-function manifold, rather than the spatial volume of a black-hole interior or an Einstein-Rosen bridge. Therefore, our result should 
not be interpreted as a direct derivation of the original holographic CV conjecture or as a statement about black-hole thermodynamics. Instead, it provides an information-geometric realization of the CV idea at the level of quantum-state geometry. In this sense, the relation $V_t=2\pi K_O$ may be viewed as a state-space analogue of the holographic complexity-volume relation. This viewpoint is complementary to recent studies connecting Krylov complexity with JT gravity, double-scaled SYK models and the growth of two-sided wormholes in AdS$_2$ \cite{Rabinovici:2023yex,Xu:2024gfm}. A more direct connection with black-hole thermodynamics, bulk gravitational dynamics or other quantum-gravity 
conjectures remains beyond the scope of the present work.

The present result also admits alternative interpretations. Rather than being viewed only 
as a generalized CV conjecture, the equality $V_t=2\pi K_O$ can be understood as a geometric encoding of Krylov-space operator growth in the two-mode squeezed-state manifold. This perspective may be useful in several related areas. In quantum information, two-mode squeezed states and Fubini-Study geometry are closely related to state 
preparation, parameter estimation and Gaussian quantum protocols. In quantum field theory, Krylov complexity provides a diagnostic of operator growth and state complexity. In quantum chaos, the Lanczos coefficients and the growth of Krylov complexity are 
connected to scrambling and operator spreading. Moreover, since information-geometric quantities and complexity measures can be sensitive to changes in the structure of quantum states, similar ideas may be useful for diagnosing quantum phase transitions. These directions, however, require further model-dependent investigations.

Several extensions and limitations of the present work should be emphasized. First, in multimode systems, the parameter space of the Fubini-Study metric becomes higher-dimensional and may encode more complicated correlations among different modes. In this case, the direct equality between Krylov complexity and the Fubini-Study volume may require additional ingredients, such as mode entanglement, squeezing distributions and higher-order interactions. Second, for the open two-mode squeezed state, the dissipative coefficient $u_2$ encodes the noise or environmental effects. Our calculation shows that the relation $V_t=2\pi K_O$ remains valid in the strongly and weakly dissipative regime. Third, Krylov complexity should be regarded as complementary to other complexity measures. Circuit complexity depends on the choice of gate set and path in Hilbert space, while spread complexity, participation ratios and entanglement-based measures characterize different aspects of operator delocalization, correlations and scrambling. The relation $V_t=2\pi K_O$ provides a geometric interpretation of Krylov-space operator growth. We also emphasize that the present evidence is analytic rather than numerical. The equality $V_t=2\pi K_O$ is derived explicitly for the closed and open two-mode Hermitian systems considered in this work. Numerical investigations would be especially useful for testing whether similar relations persist in multimode systems, strongly interacting models, or non-Hermitian dynamics, where an analytic construction of the Krylov wave function may not be available. We leave such numerical studies for future work.

Although the present work is mainly theoretical, this relation may be tested indirectly in controllable quantum platforms. Two-mode squeezed states can be realized in quantum optical systems, trapped ions and superconducting circuits. Since our construction is based on the squeezing parameters $(r,\varphi)$, the corresponding Fubini-Study metric could in principle be reconstructed through state tomography or parameter-estimation protocols. Meanwhile, Krylov complexity may be extracted from Lanczos coefficients obtained from dynamical correlation functions or operator evolution. Tunable dissipative platforms may therefore provide a possible route to test the robustness of the generalized CV relation in weakly open systems.

\appendix

\section{The calculation of the metric $ds^2$ within the open system}
\label{appendix the calculation of the metric within the open system}
The two-mode squeezed state within the open system is written as
\begin{equation}
    |z\rangle=\frac{{\rm sech}\ r}{1+u_2\tanh r}\sum_{n=0}^{\infty}|1-u_1^2|^{\frac{n}{2}}\Bigl(\frac{-e^{2i\phi}\tanh r}{1+u_2\tanh r}\Bigr)^n|n_{\vec{k}};n_{-\vec{k}}\rangle.
\end{equation}
Next, we calculate the inner product of them as
\begin{equation}
    \langle z|z\rangle=\Bigl(\frac{{\rm sech}\ r}{1+u_2\tanh r}\Bigr)^2\frac{1}{1-A},
\end{equation}
where $A = |1 - u_1^2|\tanh^2 r / (1 + u_2 \tanh r)^2$ is utilized in the subsequent calculation for simplification. The following items that we need to be aware of
\begin{equation}
    \langle z|n|z\rangle=\Bigl(\frac{{\rm sech}\ r}{1+u_2\tanh r}\Bigr)^2\frac{1}{(1-A)^2},
\end{equation}
and another one as
\begin{equation}
    \langle z|n^2|z\rangle=\Bigl(\frac{{\rm sech}\ r}{1+u_2\tanh r}\Bigr)^2\frac{A+A^2}{(1-A)^3}.
\end{equation}
Thirdly, the term $|dz\rangle$ is written as
\begin{eqnarray}
    |dz\rangle=2ind\phi|z\rangle+\Bigl(-\frac{\sinh 2r+2u_2\cosh^2r}{2\cosh^2r(1+u_2\tanh r)}+\frac{2n}{\sinh 2r}-\frac{nu_2{\rm sech^2}r}{1+u_2\tanh r}\Bigr)dr|z\rangle.
\end{eqnarray}
Finally, we obtain the information metric within the Hermitian open system as
\begin{eqnarray}
    ds^2&=&\ \Bigl[\Bigl (\frac{2}{\sinh 2r}-\frac{u_2{\rm sech^2}r}{1+u_2\tanh r}\Bigr )^2dr^2+4d\phi^2\Bigr]\Bigl(\frac{\langle z|n^2|z\rangle}{\langle z|z\rangle}-\frac{\langle z|n|z\rangle^2}{\langle z|z\rangle^2}\Bigr)\nonumber\\
    &=&\ \Bigl[\Bigl (\frac{2}{\sinh 2r}-\frac{u_2{\rm sech^2}r}{1+u_2\tanh r}\Bigr )^2dr^2+4d\phi^2\Bigr]\frac{2A}{(1-A)^2}.
\end{eqnarray}
Using the definition
\begin{equation}
A(r)=\frac{|1-u_1^2|\tanh^2 r}{(1+u_2\tanh r)^2},
\end{equation}
the factor appearing in the metric can be rewritten as
\begin{equation}
\frac{A}{1-A}
=
\frac{|1-u_1^2|\tanh^2 r}
{1+2u_2\tanh r+(u_2^2-|1-u_1^2|)\tanh^2 r}.
\end{equation}
For the positive branch of the volume element, one has
\begin{equation}
\sqrt{g}\,dr\,d\varphi
=
\left|d\left(\frac{A}{1-A}\right)\right|d\varphi . .
\end{equation}
Therefore,
\begin{equation}
V_t
=
\int_0^{2\pi}d\varphi\int_0^t dr\,\sqrt{g}
=
2\pi\left.\frac{A}{1-A}\right|_{r=t}
=
2\pi
\frac{|1-u_1^2|\tanh^2 t}
{1+2u_2\tanh t+(u_2^2-|1-u_1^2|)\tanh^2 t}
=
2\pi K_O .
\end{equation}
This shows explicitly how the Fubini-Study volume reduces to the Krylov complexity for the open two-mode squeezed state.

\acknowledgments

LHL and KHZ are funded by National Natural Science Foundation of China (NSFC) with grant NO. 12165009, Hunan Natural Science Foundation with grant NO. 2023JJ30487 and NO. 2022JJ40340. HQZ is funded by NSFC with grant NO. 12175008.



\bibliography{microlensing_milky_way}

\bibliographystyle{JHEP}
\end{document}